%% file: main.tex
\documentclass{article}
\usepackage{spconf,amsmath,graphicx}
\usepackage{authblk}

\include{macro}

\title{Image Reconstruction Without Explicit Priors}
\name{Angela F. Gao$^{1*}$, Oscar Leong$^{1*}$, He Sun$^2$, Katherine L. Bouman$^1$}
\address{$^1$Computing and Mathematical Sciences, California Institute of Technology,  $^2$Peking University
\newline $^*$ denotes equal contribution}
\begin{document}
%
\setlength{\abovedisplayskip}{6pt}
\setlength{\belowdisplayskip}{6pt}
 \setlength{\textfloatsep}{10pt }
 \setlength{\abovecaptionskip}{1pt} 
\maketitle
\begin{abstract}

We consider solving ill-posed imaging inverse problems without access to an explicit image prior or ground-truth examples. An overarching challenge in inverse problems is that there are many undesired images that fit to the observed measurements, thus requiring image priors to constrain the space of possible solutions to more plausible reconstructions. However, in many applications it is difficult or potentially impossible to obtain ground-truth images to learn an image prior. Thus, inaccurate priors are often used, which inevitably result in biased solutions. Rather than solving an inverse problem using priors that encode the explicit structure of any one image, we propose to solve a set of inverse problems jointly by incorporating prior constraints on the \textit{collective structure} of the underlying images.The key assumption of our work is that the ground-truth images we aim to reconstruct share common, low-dimensional structure. We show that such a set of inverse problems can be solved simultaneously by learning a shared image generator with a low-dimensional latent space.
The parameters of the generator and latent embedding are learned by maximizing a proxy for the Evidence Lower Bound (ELBO). Once learned, the generator and latent embeddings can be combined to provide reconstructions for each inverse problem.
The framework we propose can handle general forward model corruptions, and we show that measurements derived from only a few ground-truth images ($O(10)$) are sufficient for image reconstruction without explicit priors. 

\end{abstract}

\input{sections/sec_intro}
\input{sections/sec_approach}

\input{sections/sec_results_final}
\input{sections/sec_conclusion}

\bibliographystyle{IEEEbib}
\bibliography{references.bib}
\end{document}

%% file: macro.tex
\usepackage{latexsym,amssymb,amsmath,amsthm,graphics,graphicx,float,psfrag, epsfig, color, authblk, enumerate}

\usepackage{url,amssymb,amsmath,amsthm,amscd,paralist,bbm,wrapfig}
\usepackage{bm}
\usepackage{algorithm}
\usepackage{algorithmic} 

\usepackage{mathtools}

\usepackage[numbers]{natbib}

\newcommand{\E}{\mathbb{E}}

\DeclareMathOperator*{\KL}{KL}
\DeclareMathOperator*{\ELBO}{ELBO}
\DeclareMathOperator*{\ELBOProxy}{ELBOProxy}

\DeclarePairedDelimiterX{\infdivx}[2]{(}{)}{%
  #1\;\delimsize\|\;#2%
}
\newcommand{\infdiv}{D_{\KL}\infdivx}


%% file: sections/sec_intro.tex
\vspace{-4mm}
\section{Introduction}
\vspace{-3mm}
In imaging inverse problems, the goal is to recover the ground-truth image from corrupted measurements, where the measurements and image are related via a forward model: $y = f(x) + \eta$. Here, $y$ are our measurements, $x$ is the ground-truth image, $f$ is a known forward model, and $\eta$ is noise. Such problems are ubiquitous and include denoising \cite{Burger2012}, super-resolution \cite{CandesGranda}, compressed sensing \cite{CandesRombergTao2006, Dono2007}, and phase retrieval \cite{Numerics-of-PR}. These problems are often ill-posed: there are many images that are consistent with the observed measurements. Thus, traditionally structural assumptions on the image are required to reduce the space of possible solutions. We encode these structural assumptions in an \textit{image generation model (IGM)}, which could take the form of either an implicit or explicit image prior. 

In order to define an IGM it is necessary to have knowledge of the structure of the underlying image distribution. If ground-truth images are available, then an IGM can be learned directly \cite{venkatakrishnan2013pnp, Boraetal17}. However, this approach requires access to an abundance of clean data and there are many scientific applications (e.g., geophysical imaging and astronomical imaging) where we do not have access to ground-truth images. Collecting ground-truth images in these domains can be extremely invasive, time-consuming, expensive, or even impossible. For instance, how should we define an IGM for black hole imaging without having ever seen a direct image of a black hole or knowing what one should look like?
Moreover, classical approaches that utilize hand-crafted IGMs \cite{TV-ROF} are prone to human bias \cite{akiyama2019first4}.

In this work, we show how one can solve a set of ill-posed image reconstruction tasks without access to an explicit image prior. The key insight of our work is that knowledge of common structure across multiple problems is sufficient regularization alone. In particular, we suppose we have access to a collection of measurements $\{y^{(i)}\}_{i=1}^N$ which are observed through a forward model $y^{(i)} := f(x^{(i)}) + \eta^{(i)}$.  An important assumption we make is that the ground-truth images $\{x^{(i)}\}_{i=1}^N$ are drawn from the same distribution (unknown {\it a priori}) and share common, low-dimensional structure. This assumption is satisfied in a number of applications where ground-truth images are not available. For instance, although we might not know what a black hole looks like, we might expect it to be similar in appearance over time. Our main result is that one can exploit this common structure by jointly inferring 1) a single generator $G_{\theta}$ and 2) $N$ low-dimensional latent distributions $q_{\phi^{(i)}}$, such that the distribution induced by the push-forward of $q_{\phi^{(i)}}$ through $G_{\theta}$ approximates the posterior $p(x| y^{(i)})$ for each example $i \in [N]$. 




%% file: sections/sec_approach.tex
\vspace{-4mm}
\section{Approach}
\vspace{-2mm}

\begin{figure}
    \centering
    \includegraphics[width=0.48\textwidth]{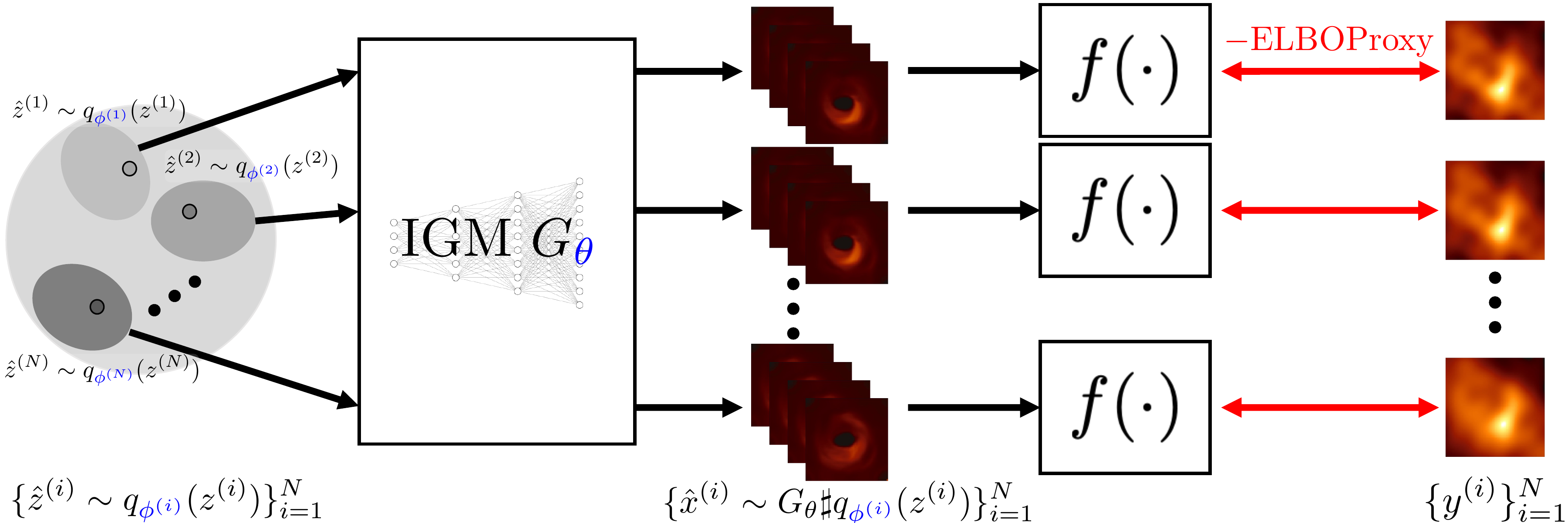}
    \caption{
    We use a set of $N$ measurements $\{y^{(i)}\}_{i=1}^N$ from $N$ different ground-truth images to infer  $\{\phi^{(i)}\}_{i=1}^N$, the parameters of the latent distributions, and $\theta$, the parameters of the generative network $G$. All inferred parameters are colored as blue and the loss is given by Equation \eqref{eqref:learning-objective}. Here, $G\sharp P$ denotes the push-forward of distribution $P$ induced by $G$.}
    \label{fig:arch}
\end{figure}
\vspace{-2mm}

In this work, we propose to solve a set of inverse problems without access to an IGM by assuming that the set of ground-truth images have common, low-dimensional structure. Other works have considered solving linear inverse problems without an IGM \cite{Lehtinenetal18, liu2020rare}, but assume that one has access to multiple independent measurements of a single, static ground-truth image, limiting their applicability to many real-world problems. In contrast, in our work we do not require multiple measurements of the same ground-truth image.

\vspace{-2mm}

\subsection{Motivation for ELBO as a model selection criterion} \label{sec:ELBO-intro}
\vspace{-1mm}

In order to accurately learn an IGM, we motivate the use of the Evidence Lower Bound (ELBO) as a loss by showing that it provides a good criterion for selecting a prior model. Suppose we are given noisy measurements from a single image: $y = f(x) + \eta$. In order to reconstruct the image $x$, we traditionally first require an IGM $G$ that captures the distribution $x$ was sampled from. A natural approach would be to find or select the model $G$ that maximizes the model posterior distribution $ p(G |y) \propto p(y | G)p(G).$
That is, conditioned on the noisy measurements, find the model of highest likelihood. Unfortunately computing $p(y | G)$ is intractable, but we show that it can be well approximated using the ELBO.

To motivate our discussion, consider estimating the conditional posterior $p(x|y,G)$ by learning the parameters $\phi$ of a variational distribution $q_{\phi}(x)$. Observe that the definition of the KL-divergence followed by using Bayes' theorem gives \begin{align*}
    & \infdiv{q_{\phi}(x)}{p(x|y,G)} 
     = \log p(y|G) \\
     & - \mathbb{E}_{x \sim q_{\phi}(x)}[\log p(y|x,G) + \log p(x|G) - \log q_{\phi}(x)].
\end{align*} The ELBO of an IGM $G$ given measurements $y$ is defined by \begin{align}
    \ELBO(G,q_{\phi};y) & := \mathbb{E}_{x \sim q_{\phi}(x)}[\log p(y|x,G) + \log p(x|G) \nonumber \\
    & - \log q_{\phi}(x)].
    \label{eq:ELBO} 
\end{align} Rearranging the above equation and using the non-negativity of the KL-divergence, we see that we can lower bound the model posterior as \begin{align}
    \log p(G|y) \geqslant \ELBO(G, q_{\phi};y) + \log p(G) - \log p(y).
\end{align} Note that $-\log p(y)$ is independent of the parameters of interest, $\phi$. If the variational distribution $q_{\phi}(x)$ is a good approximation to the posterior $p(x|y,G)$, $D_{\mathrm{KL}} \approx 0$ so maximizing $\log p(G|y)$ with respect to $G$ is approximately equivalent to maximizing $\ELBO(G, q_{\phi};y) + \log p(G)$.

 Each term in the ELBO objective encourages certain properties of the IGM $G$. In particular, the first term, $\E_{x \sim q_{\phi}(x)}[\log p(y|x,G)]$, requires that $G$ should lead to an estimate that is consistent with our measurements $y$. 
 The second term, $\E_{x \sim q_{\phi}(x)}[\log p(x|G)]$, encourages images sampled from $q_{\phi}(x)$ to have high likelihood under our model $G$. 
 The final term is the entropy term, $\E_{x \sim q_{\phi}(x)} [-\log q_{\phi}(x)]$, which encourages a $G$ that leads to ``fatter'' minima that are less sensitive to small changes in likely images $x$ under $G$.

\vspace{-4mm}

\subsection{ELBOProxy}
\vspace{-1mm}
Some IGM are explicit (e.g., Gaussian image prior), which allows for direct computation of $\log p(x|G)$. In this case, we can optimize the ELBO defined in Equation~\eqref{eq:ELBO} directly and then perform model selection. However, an important class of IGMs that we are interested in are those given by deep generative networks. Such IGMs are not probabilistic in the usual Bayesian interpretation of a prior, but instead implicitly enforce structure in the data. Moreover, terms such as $\log p(x|G)$ can only be computed directly if we have an injective map \cite{kothari2021trumpets}. This architectural requirement limits the expressivity of the network. Hence, we instead consider a proxy of the ELBO that is especially helpful for deep generative networks. Suppose our image generation model is of the form $x = G(z)$ where $G$ is a generative network and $z$ is a latent vector. Introducing a variational family for our latent representations $z \sim q_{\phi}(z)$ and using $\log p(z| G)$ in place of $\log p(x|G)$, we arrive at the following proxy of the ELBO:
\begin{align}
\ELBOProxy(G, q_{\phi};y) & := \E_{z \sim q_{\phi}(z)}[\log p(y |G(z)) \nonumber\\
& + \log p(z | G) - \log q_{\phi}(z)]. \label{eqref:ELBOProxy}  
\end{align} When $G$ is injective and $q_{\phi}(x)$ is the push-forward of $G$ through $q_{\phi}(z)$, then this proxy is exactly the ELBO in Eq. \eqref{eq:ELBO}. While $G$ may not necessarily be injective, we show empirically that the ELBOProxy is a useful criterion for selecting such models. 


\noindent \textbf{Toy example:} To illustrate the use of the $\ELBOProxy$ as a model selection criterion, we conduct the following experiment that asks whether the $\ELBOProxy$ can identify the best model from a given set of plausible IGMs. For this experiment, we use the MNIST dataset \cite{MNIST} and consider two inverse problems: denoising and phase retrieval. We train a generative model $G_{c}$ on each class $c \in \{0,1,2,\dots,9\}$. Hence, $G_{c}$ is learned to generate images from class $c$ via $G_{c}(z)$ where $z \sim \mathcal{N}(0,I)$. Then, given noisy measurements $y_{c}$ of a single image from class $c$, we ask whether the generative model $G_{c}$ from the appropriate class would achieve the best $\ELBOProxy$. For denoising, our measurements are $y_c = x_c + \eta_c$ where $ \eta_c\sim \mathcal{N}(0, \sigma^2 I)$ and $\sigma = \sqrt{0.5}$. For phase retrieval, $y_c = |\mathcal{F}(x_{c})| + \eta_c$ where $\mathcal{F}$ is the Fourier transform and $\eta_{c} \sim \mathcal{N}(0, \sigma^2 I)$ with $\sigma = \sqrt{0.05}$.

\begin{figure}
    \centering
    \includegraphics[width=0.48\textwidth]{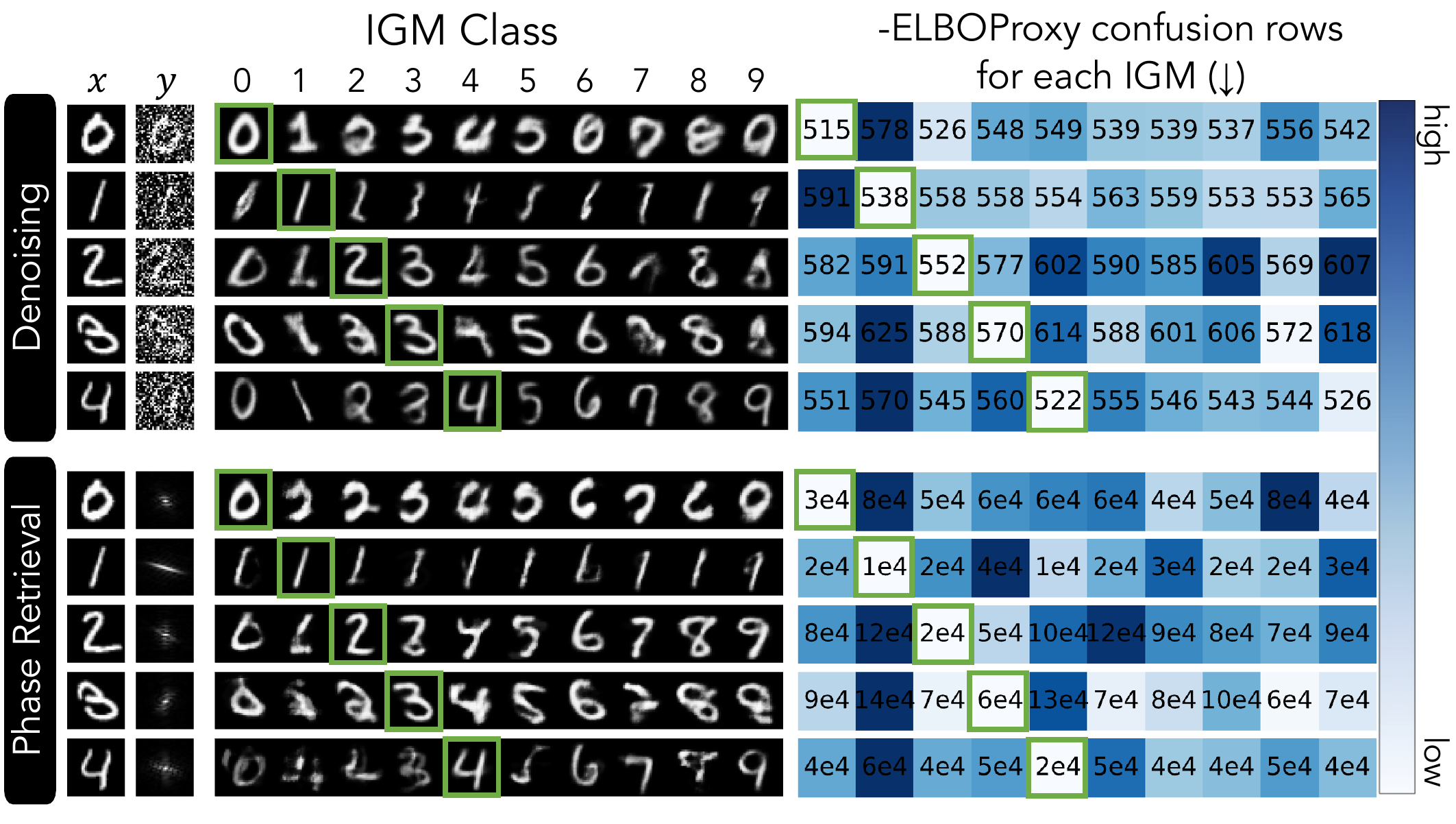}
    \caption{We consider two inverse problems: denoising and phase retrieval. Left: the two leftmost columns correspond to the ground truth image $x_c$ and the noisy measurements $y_c$. Center: in each row, we show the means of the distribution induced by the push-forward of $G_j$ and each latent distribution $z \sim q_{\phi_j}$ for $j \in \{0,\dots,9\}$. 
    Right: each row of the array corresponds to the $-\ELBOProxy$ achieved by each model in reconstructing the images. Here, lower is better. Boxes highlighted in green correspond to the best $-\ELBOProxy$ values in each row. In all cases, the correct model was chosen.  
    }
    \label{fig:MNIST_ELBO_exp}
\end{figure}

We construct $5 \times 10$ arrays for each problem, where in the $i$-th row and $j$-th column, we compute the $-\ELBOProxy$ obtained by using model $G_{{j-1}}$ to reconstruct images from class $i-1$. We calculate $\ELBOProxy(G_{c}, q_{\phi_c};y_c)$ by parameterizing $q_{\phi_c}$ with a Normalizing Flow and optimizing network weights $\phi_c$ to maximize \eqref{eqref:ELBOProxy}. Results are shown in Fig.~\ref{fig:MNIST_ELBO_exp}. We note that all of the correct models are chosen in both denoising and phase retrieval. We also note some interesting cases where the $\ELBOProxy$ values are similar for certain cases, such as when recovering the $3$ or $4$ image when denoising. For example, when denoising the $4$ image, both $G_{4}$ and $G_{9}$ achieve comparable $\ELBOProxy$ values. By carefully inspecting the noisy image of the $4$, one can see that both models are reasonable given the structure of the noise. 

\vspace{-3mm}
\subsection{Learning the IGM to solve inverse problems}\label{sec:learning}

\vspace{-1mm}

As the previous section illustrates, the $\ELBOProxy$ provides a good criterion for choosing an appropriate IGM from noisy measurements. Here, we consider the task of learning the IGM $G$ directly from corrupted data. 
We consider the setting where we have access to a collection of $N$ measurements $y^{(i)} = f(x^{(i)}) + \eta^{(i)}$ for $i \in [N]$. The key assumption we make is that common, low-dimensional structure is shared across the underlying images $\{x^{(i)}\}_{i=1}^N$. 

We propose to find a \textit{shared} generator $G_{\theta}$ with weights $\theta$ along with latent distributions $q_{\phi^{(i)}}$ that can be used to reconstruct the full posterior of each image $x^{(i)}$ from its corresponding noisy measurements $y^{(i)}$. This approach is illustrated in Fig.~\ref{fig:arch}. Having the generator be shared across all images helps capture their common collective structure. Each corruption, however, could induce its own complicated image posteriors. Hence, we assign each measurement $y^{(i)}$ its own latent distribution to capture the differences in their posteriors. While the learned distribution may not necessarily be the true image posterior (as we are optimizing a proxy of the ELBO), it still captures a distribution of images that fit to the observed measurements.


 More explicitly, given a set of measurements $\{y^{(i)}\}_{i = 1}^N$, we optimize the $\ELBOProxy$ from Equation \eqref{eqref:ELBOProxy} to jointly infer a  generator $G_{\theta}$ and variational distributions $\{q_{\phi^{(i)}}\}_{i = 1}^N$:
 \vspace{-2mm}
\begin{align}
 \max_{\theta, \phi^{(i)}} & \frac{1}{N}\sum_{i = 1}^N \ELBOProxy(G_{\theta},q_{\phi^{(i)}}; y^{(i)}) + \log p(G_{\theta}). \label{eqref:learning-objective} 
\end{align}
The expectation in this objective is approximated via Monte Carlo sampling. In terms of choices for $\log p(G_{\theta})$, we can add additional regularization to promote particular structures, such as smoothness. Here, we consider having sparse neural network weights as a form of implicit regularization and use dropout during training to represent $\log p(G_{\theta})$ \cite{srivastava2014dropout}.
 
 Once a generator $G_{\theta}$ and variational parameters $\{\phi^{(i)}\}_{i=1}^N$ have been learned, we solve the $i$-th inverse problem by simply sampling $\hat{x}^{(i)} = G_{\theta}(\hat{z}^{(i)})$ where $\hat{z}^{(i)} \sim q_{\phi^{(i)}}$. Producing multiple samples for each inverse problem can help visualize the range of uncertainty under the learned IGM $G_{\theta}$, while taking the average of these samples empirically provides clearer estimates with better metrics in terms of PSNR or MSE. We report PSNR outputs in our subsequent experiments. 

%% file: sections/sec_results_final.tex
\vspace{-4mm}
\section{Experimental Results}
\vspace{-3mm}


We now consider solving a set of inverse problems via the framework described in \ref{sec:learning}. For each of these experiments, we use a multivariate Gaussian distribution to parameterize each posterior distribution $q_{\phi^{(i)}}$, and we use a Deep Decoder \cite{HH2018} with $6$ layers of size $150$, a latent size of $40$, and a dropout of $10^{-4}$ as the IGM. Each multivariate Gaussian distribution is parameterized by a mean and covariance matrix $\{\mu^{(i)},L^{(i)}L^{{(i)}^T} + \varepsilon I\}_{i=1}^N$, with $\varepsilon = 10^{-3}$ is added to the covariance matrix to help with stability of the optimization.

\begin{figure}[ht]
    \centering
    \includegraphics[width=0.48\textwidth]{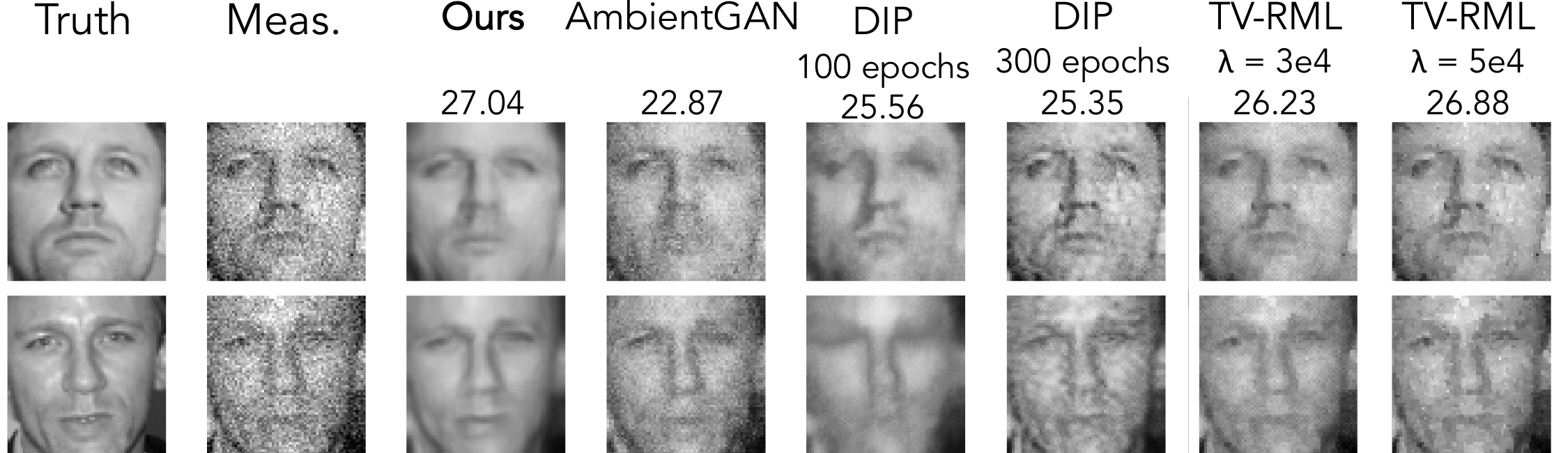}
    \caption{\textbf{Denoising baseline comparisons.} We compare to AmbientGAN, DIP, and TV-RML with weight $\lambda$ and report the average PSNR across all 95 reconstructions. We show both early stopping and full training results using DIP. Our results exhibit higher PSNR than all other baselines while maintaining distinct features of the ground-truth images.}
    \label{fig:bondbaselines}
\end{figure}

\vspace{1mm}
\noindent \textbf{Denoising:} We show results on denoising noisy images of a single face from the PubFig \cite{kumar2009attribute} dataset in Fig.~\ref{fig:bondbaselines}. The measurements are defined by $y = x + \eta$ where $\eta \sim \mathcal{N}(0,\sigma^2I)$ with an SNR of $\sim$15 dB. Our method is able to remove much of the noise and recovers small scale features, even with only 95 examples. Our reconstructions also substantially outperform the baseline methods AmbientGAN \cite{bora2018ambientgan}, Deep Image Prior (DIP) \cite{ulyanov2018deep}, and regularized maximum likelihood using total variation (TV-RML), as shown in Fig.~\ref{fig:bondbaselines}. Unlike DIP, our method does not seem to overfit and does not require early stopping. Our method also does not exhibit noisy artifacts like those seen in AmbientGAN results.

\begin{figure}[ht]
    \centering
    \includegraphics[width=.48\textwidth]{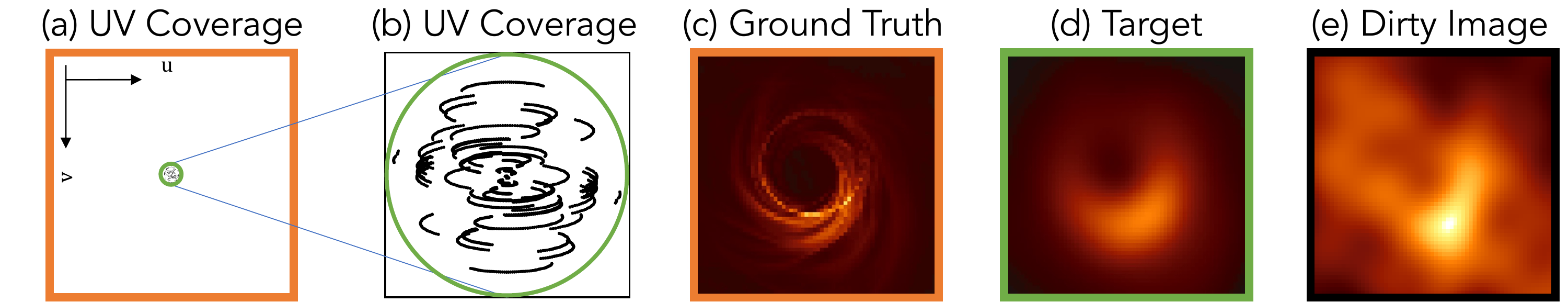}
    \caption{\textbf{Visualization of the intrinsic resolution of the EHT compressed sensing measurements.} The EHT measures sparse spatial frequencies of the image (i.e., components of the image's Fourier transform). In order to generate the ground truth image (c), all frequencies in the entire domain of (a) must be used. Restricting spatial frequencies to the ones in (a) and (b)'s green circle generates the target (d). 
    The EHT samples a subset of this region, indicated by the sparse black samples in (b). Naively recovering an image using only these frequencies results in the \textit{dirty image} (e), which is computed by $A^Hy$. 
    The 2D spatial Fourier frequency coverage represented with $(u, v)$ positions is referred to as the UV coverage.
    }
    \label{fig:uvcoverage}
\end{figure}

\vspace{1mm}
\noindent \textbf{Compressed sensing:} We consider a compressed sensing problem inspired by astronomical imaging of black holes with the Event Horizon Telescope (EHT): suppose we are given access to measurements of the form $y = Ax + \eta,\ \eta \sim \mathcal{N}(0,\sigma^2 I)$ where $A\in \mathbb{C}^{p \times n}$ is a low-rank compressed sensing matrix arising from interferometric telescope measurements. This problem is ill-posed and requires the use of priors or regularizers to recover a reasonable image \cite{akiyama2019first4}. Moreover, it is impossible to acquire ground-truth images of black holes, so any explicit prior defined {\it a priori} will exhibit human bias. However, we can assume that although the black hole evolves, it will not change drastically from day-to-day. 

\begin{figure}[ht]
    \centering
    \includegraphics[width=0.4\textwidth]{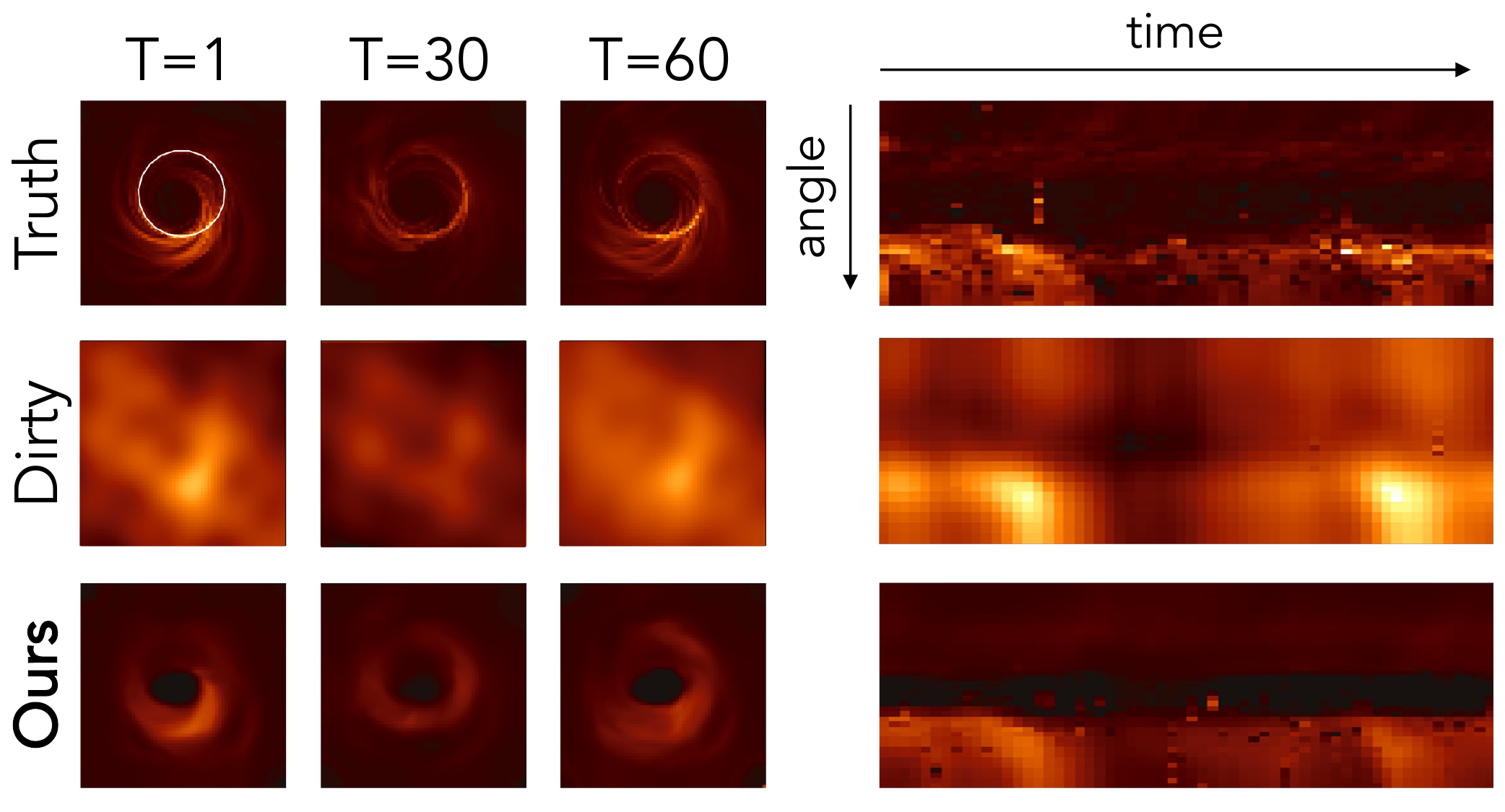}
    \caption{\textbf{Compressed sensing with a video of a black hole.} We demonstrate our method described in Section~\ref{sec:learning} using 60 images from an evolving black hole target. Here we show the ground-truth, dirty ($A^Hy$, see Fig. \ref{fig:uvcoverage}), and mean reconstruction images, respectively. We also show the unwrapped space × time image, which is taken along the overlaid white ring illustrated in the T=1 ground-truth image. The bright-spot's temporal trajectory of our reconstruction matches that of the ground-truth.}\label{fig:BH_compressed_sensing}
\end{figure}

We show results on 60 frames from a video of a simulated evolving black hole target \cite{porth2019event, akiyama2019first5} with an SNR of $\sim$32 dB in Fig.~\ref{fig:BH_compressed_sensing}. The reference ``target image'' is the ground-truth filtered with a low-pass filter that represents the maximum resolution intrinsic to the EHT array (visualized and explained in Fig.~\ref{fig:uvcoverage}). 
Our method is not only able to reconstruct the large scale features of the ground-truth images without any aliasing artifacts, but also  achieves a level of super-resolution.

\noindent \textbf{Non-Convex Phase Retrieval:} Here we demonstrate our approach on solving non-convex phase retrieval. Our measurements are described by $y = |\mathcal{F}(x)| + \eta$ where $\mathcal{F}(\cdot)$ is a complex-valued linear operator (either Fourier or undersampled Gaussian measurements) and $\eta \sim \mathcal{N}(0, \sigma^2 I)$. 

\begin{figure}[h!]
    \centering
    \includegraphics[width=0.35\textwidth]{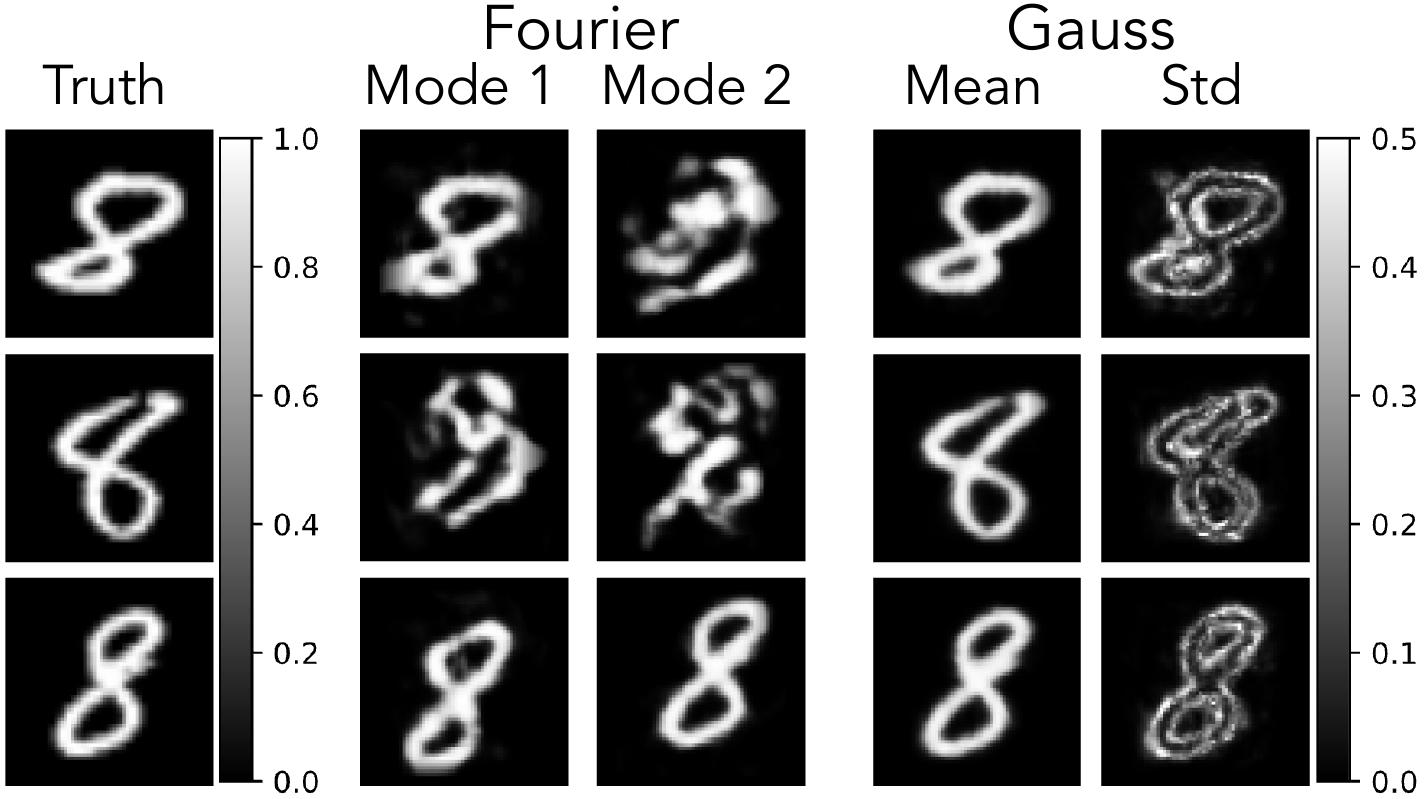}
    \caption{\textbf{Phase retrieval with MNIST 8's.} We shown an example of our method applied to the challenging non-convex problems of Fourier and complex Gaussian phase retrieval. 
    }
    \label{fig:Mnist_phase_retrieval}
\end{figure}

Fig.~\ref{fig:Mnist_phase_retrieval} shows results from a set of $N = 150$ measurement examples of the MNIST 8's class with an SNR of $\sim$52 dB. The true posterior is multi-modal due to the phase ambiguity and non-convexity of the problem. Our reconstructions have features similar to the digit $8$, but contain artifacts in the Fourier case. These artifacts are only present in the Fourier case due to additional ambiguities (i.e., flipping and spatial-shifts), which lead to a more complex posterior \cite{SparsePRoverview}.

%% file: sections/sec_conclusion.tex
\vspace{-4mm}
\section{Conclusion}
\vspace{-2mm}


In this work, we showcased how one can solve a set of inverse problems without an IGM by leveraging common structure present across the underlying ground-truth images. We demonstrated that even with a small set of corrupted measurements, one can jointly solve these inverse problems by directly learning
an IGM that maximizes a proxy of the ELBO.
Overall, our work showcases the possibilities of solving inverse problems in a ``prior-free'' fashion, free from human bias typical of ill-posed image reconstruction. We believe our approach can aid automatic discovery of novel structure from scientific measurements without access to clean data, leading to potentially new avenues for scientific discovery.
\vspace{1mm}

\noindent \textbf{Acknowledgements}
This work was sponsored by NSF Award 2048237 and 1935980, an Amazon AI4Science Partnership Discovery Grant, and the Caltech/JPL President’s and Director’s Research and Development Fund  (PDRDF).  This research was carried out at the Jet Propulsion Laboratory and Caltech under a contract with the National Aeronautics and Space Administration and funded through the PDRDF.